\documentclass{amsart}
\usepackage{amsmath}
\usepackage{amssymb}
\usepackage{amsthm}
\usepackage{color}
\usepackage{array}
\usepackage{caption}
\usepackage{graphicx, subfigure}
\usepackage{cite}
\usepackage{mathrsfs}
\usepackage{algorithm}
\usepackage{algpseudocode}
\usepackage{multirow}
\usepackage{epstopdf}
\ifpdf
  \DeclareGraphicsExtensions{.png,.pdf,.png,.jpg}
\else
  \DeclareGraphicsExtensions{.png}
\fi

\title[Iterative splitting method for Heston model]{An iterative splitting method for pricing European options under the Heston model}
\author[H.S. Li, Z.Y. Huang] {Hongshan Li, Zhongyi Huang$^\dag$}

\address[H.S. Li, Z.Y. Huang]{Dept. of Mathematical Sciences, Tsinghua Univ., Beijing 100084, China}
\email{lhsusan1016@163.com; zhongyih@tsinghua.edu.cn}
\thanks{This work was partially supported by NSFC Project No.~11871298.}
\thanks{$^\dag$ Corresponding author.}

\date{\today}

\begin{document}

\maketitle

\begin{abstract}
In this paper, we propose an iterative splitting method to solve the partial differential equations in option pricing problems. We focus on the Heston stochastic volatility model and the derived two-dimensional partial differential equation (PDE). We take the European option as an example and conduct numerical experiments using different boundary conditions. The iterative splitting method transforms the two-dimensional equation into two quasi one-dimensional equations with the variable on the other dimension fixed, which helps to lower the computational cost. Numerical results show that the iterative splitting method together with an artificial boundary condition (ABC) based on the method by Li and Huang (2019) gives the most accurate option price and Greeks compared to the classic finite difference method with the commonly-used boundary conditions in Heston (1993).
\end{abstract}

\section{Introduction}
An option is a financial instrument that provides the investor with the opportunity to buy or sell an asset at a specified price. Options make it possible for investors to  benefit from the return of the underlying asset at a low cost and play an important role in hedging risk as well.

Research on option pricing can be traced back to Bachelier's work in 1900 \cite{bachelier1900theorie}, since then an increasing number of mathematical tools have been introduced into the financial field \cite{merton1998applications}. In 1973, the seminal work of Black and Scholes opened a new era of the option pricing theory \cite{black1973pricing}. After that, the option market expanded rapidly and option pricing attracted considerable attention of researchers.

Due to the volatility smile and skew reflected by the market data,  the assumption of constant volatility in the Black-Scholes model need relaxing and different models have been proposed to better reflect the market features based on the Black-Scholes model. Assume that the volatility of the underlying is a function of the price and time, then we get to the well-accepted local volatility model  \cite{dupire1994pricing} \cite{derman1994riding}. By calibrating it to the market prices of European options, we can obtain the implied volatility surface which can be used to price other options. But it always leads to the behavior of the volatility smile opposite to the market as the price changes\cite{hagan2002managing}. In jump diffusion models, Poisson processes are added into the dynamic of the underlying to capture the phenomena of the volatility \cite{merton1976option,duffie2000transform,kou2002jump}. However, they are not applicable to riskless hedging. L\'evy processes are continuous-time stochastic processes with stationary independent increments and are used to model the return of the underlying \cite{chan1999pricing,geman2001time,carr2003stochastic}. The Black-Scholes model and jump diffusion models can be seen as special cases of models based on L\'evy processes. Because of the inherent property of L\'evy processes, i.e., independent increments, they are not able to capture the dependence structure. Stochastic volatility models are constructed with an additional stochastic process to describe the dynamic of the volatility \cite{johnson1987option,wiggins1987option,hull1987pricing,stein1991stock,Heston1993,fouque2000mean}. Among them, Heston's stochastic volatility model is the most popular and widely used in both academia and industry. The volatility in the Heston model follows the Cox-Ingersoll-Ross (CIR) process, which ensures that it cannot be negative. The dependence structure can be incorporated and it has a closed-form solution when pricing European options.

As we have shown above, different continuous-time stochastic processes are introduced to model the price or other state variables such as the volatility of the underlying and always appear in literature as stochastic differential equations (SDEs). Under the risk neutral probability measure, the price of the option can be represented as the conditional expectation of the discounted payoff determined by the contract. By the Feynman-Kac theorem, it can be considered as a solution of a partial differential equation (PDE) corresponding to the dynamics or SDEs we assume. Closed-form solutions are difficult to derived or calculated for some of the models and options so that a variety of numerical methods have been developed. By lattice methods, the continuous distribution is transformed into a discrete distribution, which is much easier to understand and implement \cite{cox1979option,boyle1988lattice}. However, lattice methods are not suitable for complicated models. To deal with the continuous-time stochastic processes or SDEs directly, we have the well-known Monte Carlo simulation method \cite{boyle1977options,johnson1987option,boyle1997monte}. It is flexible and applies to high-dimensional problems, but it gives solutions with probable errors and requires a large number of simulation paths to guarantee the accuracy. Researchers have tried to use numerical integration methods to compute the price of the option based on the Fourier Transform with the characteristic functions of the distributions of the underlying \cite{carr1999option,fang2008novel}. These methods are efficient and can be useful for calibration, while they always need careful derivation of the formulae for different models and options. Another class of methods is to price options by solving the PDEs and a wide range of numerical techniques are then available including finite difference methods \cite{brennan1978finite,hull1990valuing,zvan2000pde,ikonen2004operator,foulon2010adi}, finite element methods \cite{forsyth1999finite,winkler2001valuation} and so on. PDE methods especially finite difference methods are intuitive and interpretable, but memory space and computational time they cost may increase rapidly as the dimension of the problem grows.

In this paper, we take the Heston stochastic volatility model as an example and try to give a new numerical method for the option pricing problem, which can maintain the advantages of finite difference methods and reduce the influence of the dimension of the PDE. Compared with the Black-Scholes model, the Heston model involved the CIR process to describe the volatility of the underlying, which is a mean-reverting process and can be seen as adding fluctuations to a given mean volatility to some extent. Accordingly, the PDE derived from the Heston model can be seen as adding terms related to the volatility to the PDE corresponding to the Black-Scholes model, and the dimension is then increased from one to two. On this basis, we propose an iterative splitting method to split the operators and solve the quasi one-dimensional PDEs iteratively. We make use of the analytical solution of the Black-Scholes model, lower the computational cost and maintain the simplicity, flexibility and interpretability of the PDE methods. A mixed method is also used to guarantee the convergence of the iterative process. We conduct numerical experiments to price the European options. Boundary conditions from \cite{Heston1993} and artificial boundary conditions obtained with the methods in \cite{li2019artificial} are used for a comparison. Numerical results show that the iterative splitting method together with an artificial boundary condition based on the method from \cite{li2019artificial} gives the most accurate solutions and Greeks. By contrast with a classic two-dimensional finite difference scheme, its solutions have smaller errors as well. Besides, the iterative splitting method can be easily extended to other models and options.

The paper is organized as follows. Section \ref{sec:Heston} presents the PDE derived from the Heston stochastic volatility model. The iterative splitting method is proposed in Section \ref{sec:itsp} and the corresponding algorithms are also given. In Section \ref{sec:bcs}, we derive and show different artificial boundary conditions with the methods in \cite{li2019artificial}. Section \ref{sec:num} gives the results of numerical experiments with different parameters, boundary conditions and numerical methods. We summarize in Section \ref{sec:con}.

\section{The Heston stochastic volatility model}\label{sec:Heston}
Denote the asset spot price and its variance by $S(t)$ and $v(t)$. According to \cite{Heston1993}, assume that $S(t)$ and $v(t)$ satisfy the following two stochastic processes:
\begin{align}
  \label{eq:stockSDE}
  dS(t)=\mu S(t)dt + \sqrt{v(t)}S(t)dW^{(1)}(t),\\
  \label{eq:volSDE}
  dv(t)=\kappa\left(\theta-v(t)\right)dt+\sigma\sqrt{v(t)}dW^{(2)}(t).
\end{align}
The parameter $\mu$ is the expected return of the underlying, $\kappa$ is the speed at which the variance $v(t)$ reverts to its long run mean $\theta$ and $\sigma$ is the so-called "volatility of volatility". In addition, $W^{(1)}(t)$ and $W^{(2)}(t)$ are standard Wiener processes with a constant correlation coefficient $\rho$.

The price of a financial derivative $C(S,v,t)$ can then be considered as the solution of a Garman's type PDE:
\begin{equation}
  C_t +  \frac{1}{2}vS^2C_{SS}+\rho\sigma vSC_{Sv}+\frac{1}{2}\sigma^2vC_{vv}+rSC_S+[\kappa(\theta-v)-\lambda]C_v-rC=0,
  \label{eq:HestonPDE}
\end{equation}
where $r$ is the risk-free interest rate and the price of volatility risk $\lambda(S,v,t)$ can be set to be 0.

We take the European call option as an example and denote the maturity and the strike by $T$ and $K$. Let $\tau=T-t$ represent the time to maturity and 
\[\tilde{S}=Se^{r\tau}/K,\quad U(\tilde{S},v,\tau)=C(S,v,t)e^{r\tau}/K,\]
then the PDE \eqref{eq:HestonPDE} is equivalent to
\begin{equation}
  \label{eq:HestonPDESimple}
  \begin{aligned}
	\frac{1}{2}v\tilde{S}^2U_{ \tilde{S}\tilde{S}}+\rho\sigma v\tilde{S}U_{\tilde{S}v}+\frac{1}{2}\sigma^2vU_{vv}+\kappa(\theta-v)U_v=U_ \tau, \\
   \tilde{S}\in(0, +\infty),\quad v\in(0,+\infty),\quad \tau\in(0,T].\\
  \end{aligned}
\end{equation}

Besides, to give the price of the derivative, an initial value and proper boundary conditions are also needed. Referring to \cite{Heston1993}, the conditions can be rewritten as follows with the transforms mentioned above:
\begin{eqnarray}
\label{eq:HestonIC}
U|_{\tau=0}=(\tilde{S}-1)^+,\quad \tilde{S}\in[0,+\infty),\quad  v\in[0,+\infty),\\
\label{eq:HestonBCfirst}
U|_{\tilde{S}=0}=0,\quad v\in[0,+\infty),\quad \tau\in(0,T],\\
U_{\tilde{S}} \to 1 \quad \text{as} \quad \tilde{S}\to +\infty,\quad v\in[0,+\infty),\quad \tau\in(0,T],\\
(\kappa\theta U_v-U_\tau)|_{v=0}=0,\quad \tilde{S}\in[0,+\infty),\quad \tau\in(0,T],\\
\label{eq:HestonBClast}
U\to \tilde{S} \quad \text{as} \quad v \to +\infty,\quad \tilde{S}\in[0,+\infty),\quad \tau\in(0,T].
\end{eqnarray}

\section{An iterative splitting method for the option pricing problem}\label{sec:itsp}
First, we split the operators on the left-hand side of the PDE \eqref{eq:HestonPDESimple} into two parts. Let
\begin{equation}
\mathscr{L}_1 U=\frac{1}{2}v\tilde{S}^2U_{\tilde{S}\tilde{S}},
\label{eq:operator1}
\end{equation}
\begin{equation}
\mathscr{L}_2 U=\rho\sigma v\tilde{S}U_{\tilde{S}v}+\frac{1}{2}\sigma^2vU_{vv}+\kappa(\theta-v)U_v,
\label{eq:operator2}
\end{equation}
then the PDE \eqref{eq:HestonPDESimple} can be rewritten as
\begin{equation}
U_\tau = \mathscr{L}_1 U+ \mathscr{L}_2 U.
\label{eq:HestonPDEOp}\\
\end{equation}
If $\mathscr{L}_2 U$ is known and can be denoted by  $Q(\tilde{S}, v, \tau)$, the two-dimensional problem \eqref{eq:HestonPDESimple}-\eqref{eq:HestonBClast} with a mixed spatial-derivative term can be transformed into two quasi one-dimensional problems with the same operator.
Let
\[U=U_1+U_2,\]
we get to the following two problems:
\begin{equation}
  \begin{aligned}
	\mathscr{L}_1 U_1= U_{1,\tau},\\
	U_1|_{\tau=0} = (\tilde{S} - 1)^+,\\
	U_1|_{\tilde{S}=0}=0,\\
	U_{1,\tilde{S}} \to 1 \quad \text{as} \quad \tilde{S}\to +\infty,
  \end{aligned}
  \label{eq:PDEpart1}
\end{equation}
and
\begin{equation}
  \begin{aligned}
	\mathscr{L}_1 U_2+Q=U_{2,\tau},\\
	U_2|_{\tau=0}=0,\\
	U_2|_{\tilde{S}=0}=0,\\
	U_{2,\tilde{S}} \to 0 \quad \text{as} \quad \tilde{S}\to +\infty,\\
	(U_{2,\tau}-Q)|_{v=0}=0,\\
	U_2\to 0 \quad \text{as} \quad v \to +\infty.
  \end{aligned}
  \label{eq:PDEpart2}
\end{equation}

We can easily give the solution to the problem \eqref{eq:PDEpart1} based on the Black-Scholes formula:
\begin{equation}
  U_1=\tilde{S}\cdot N\left(\frac{\ln{\tilde{S}}+v\tau/2}{\sqrt{v\tau}}\right)-N\left(\frac{\ln{\tilde{S}}-v\tau/2}{\sqrt{v\tau}}\right).
  \label{eq:solutionU1}
\end{equation}
Here, $N(x)$ represents the cumulative distribution function of the standard normal distribution.
	
	The problem \eqref{eq:PDEpart2} can be solved only if  $Q(\tilde{S},v,\tau)=\mathscr{L}_2 U$ is known, so we try to use an iterative process to find the solution. Since the operators in the original problem have been split as  above, we call it an iterative splitting method.

Choose a time step size $\Delta\tau=T/N$, and let $\tau_n=n\cdot\Delta\tau, \ n=0, ..., N$. Denote $U^n(\tilde{S}, v, \tau)=U(\tilde{S}, v, \tau_n)$ and define $U_1^n$, $U_2^n$and $Q^n$ similarly.
For the $n$-th time step, we first calculate $U_1^n$ by \eqref{eq:solutionU1}. With the known $U^{n-1}$, we compute $Q^{n-1}=\mathscr{L}_2 U^{n-1}$ and use it to approximate $Q^n$. Then we can solve the problem \eqref{eq:PDEpart2} numerically to get $U_2^n$. Since $U^n=U^n_1+U^n_2$, $U^n$ can be updated, with which we obtain a new $Q^n$. The problem \eqref{eq:PDEpart2} can be solved again, and we can recalculate $U^n_2$, $U^n$ and $Q^n$ again and again until some termination condition is satisfied. The algorithm is shown in {\bf Algorithm \ref{alg:alg1}}.
\begin{algorithm}[htb]
  \caption{}
  \label{alg:alg1}
  	\begin{algorithmic}
  	 \State $U^0=(\tilde{S}-1)^+$
  	 \For{$n=1, ..., N$}
  	  \State Compute $U_1^n$ by \eqref{eq:solutionU1}
  	  \State $U^n=U^{n-1}$
  	  \Repeat
  	      \State Compute $Q^n=\mathscr{L}_2 U^n$
  	      \State Solve $\mathscr{L}_1 U_2+Q^n=U_{2,\tau}$ to get $U_2^n$
  	      \State $U^n=U_1^n+U_2^n$
  	  \Until the change of $U^n_2$ is small enough
  	 \EndFor	
	\end{algorithmic}
\end{algorithm}

The interval $[0, T]$ has been divided into $N$ pieces. Choose $\Delta \tilde{S}=\tilde{S}_{max}/I$ and $\Delta v=v_{max}/J$, and let $\tilde{S}_i=i\cdot\Delta \tilde{S},\ i=0,...,I,\ v_j=j\cdot\Delta v,\ j=0,...,J$, then we get a uniform grid.
Denote $U_{i,j}^n=U(\tilde{S}_i,v_j,\tau_n)$ and $U_{1,i,j}^n$, $U_{2,i,j}^n$ and $Q_{i,j}^n$ are similar.

We approximate $Q^n=\mathscr{L}_2 U^n$ with an upwind scheme as follows:
\begin{equation}
  \begin{aligned}
	Q_{i,j}^n =& \rho\sigma v_j\tilde{S}_i\frac{U_{i+1,j+1}^{n}-U_{i-1,j+1}^{n}+U_{i-1,j-1}^{n}-U_{i+1,j-1}^{n}}{4\Delta\tilde{S}\Delta v}+\frac{1}{2}\sigma^2v_j\frac{U_{i,j+1}^{n}-2U_{i,j}^{n}+U_{i,j-1}^{n}}{(\Delta v)^2}\\
&+\kappa(\theta-v_j)^+\frac{U_{i,j+1}^{n}-U_{i,j}^n}{\Delta v}+\kappa(\theta-v_j)^-\frac{U_{i,j}^{n}-U_{i,j-1}^n}{\Delta v},\quad i=1, ..., I-1, \quad j=1, ..., J-1.
  \end{aligned}
  \label{eq:Qdiscrete}
\end{equation}

To compute $U^n_2$, we give an implicit central difference scheme:
\begin{equation}
\frac{1}{2}v_j\tilde{S}_i^2\frac{U_{2,i+1,j}^{n}-2U_{2,i,j}^{n}+U_{2,i-1,j}^{n}}{(\Delta\tilde{S})^2} + Q_{i,j}^n=\frac{U_{2,i,j}^{n}-U_{2,i,j}^{n-1}}{\Delta\tau},\quad i=1,...,I-1, \quad j=1, ..., J-1.
	\label{eq:solveU2}
\end{equation}

It is noteworthy that the iterative process is not always convergent with the algorithm and the finite difference scheme above. We calculate $Q^n$ based on the solution $U^n$ totally from the last iteration and it may lead to divergence when the coefficients in the expression of $Q^n$ \eqref{eq:Qdiscrete} get large. A mixed method should be used to calculate $Q^n$ to guarantee convergence.

Let
\[Q_1=\mathscr{L}_2 U_1, \quad Q_2=\mathscr{L}_2 U_2,\quad Q=Q_1+Q_2,\]
and define $Q_1^n$, $Q_2^n$, $Q^n_{1, i, j}$, $Q^n_{2, i, j}$ for $i=0,...,I, \ j=0,...,J, \ n=0,...,N$ similarly to $U^n$. To get $U_2^n$, a central difference scheme is still used to approximate $\mathscr{L}_1$ as in \eqref{eq:solveU2}. $Q^n$ can be divided into two parts including $Q^n_1$ and $Q^n_2$. We compute $Q^n_1=\mathscr{L}_2 U^n_1$ with the same discretization of $\mathscr{L}_2$ in \eqref{eq:Qdiscrete}. In fact, $Q^n_1$ can also be calculated exactly since the expression of $U^n_1$ is known. In terms of $Q^n_2$, we still discretize $\mathscr{L}_2U_2$ as in \eqref{eq:Qdiscrete}, but instead of computing with $U_2^n$ totally from the last iteration, we make the terms on the current grid point $(i, j)$ in the discretization come from the current iteration step and keep others unchanged from the last step. We present the improved algorithm in {\bf Algorithm \ref{alg:alg2}}.
\begin{algorithm}[htb]
  \caption{}
  \label{alg:alg2}
  	\begin{algorithmic}
  	 \State $U^0=(\tilde{S}-1)^+$
  	 \For{$n=1, ..., N$}
  	  \State Compute $U_1^n$ by \eqref{eq:solutionU1}
  	  \State Compute $Q^n_1=\mathscr{L}_2 U^n_1$
  	  \State $U^n_2=U^{n-1}_2$
  	  \Repeat
  	      \State Solve $\mathscr{L}_1 U_2 + \mathscr{L}_2 U_2+Q^n_1=U_{2,\tau}$ to get $U_2^n$
  	  \Until the change of $U^n_2$ is small enough
  	  \State $U^n=U_1^n+U_2^n$
  	 \EndFor	
	\end{algorithmic}
\end{algorithm}

To make it clearer, the following numerical scheme is used to get $U_2^n$:
\begin{eqnarray}\qquad
\frac{1}{2}v_j\tilde{S}_i^2\frac{U_{2,i+1,j}^{n}-2U_{2,i,j}^{n}+U_{2,i-1,j}^{n}}{(\Delta\tilde{S})^2} - \frac{1}{2}\sigma^2v_j\frac{2U_{2,i,j}^{n}}{(\Delta v)^2} -\kappa|\theta-v_j|\frac{U_{2,i,j}^{n}}{\Delta v}+Q_{1,i,j}^n+\hat{Q}_{2,i,j}^n=\frac{U_{2,i,j}^{n}-U_{2,i,j}^{n-1}}{\Delta\tau},\\
	\label{eq:solveU22}
i=1,...,I-1, \quad j=1, ..., J-1. \nonumber
\end{eqnarray}
Here, $\hat{Q}_{2,i,j}^n$ is computed by the following expression but note that $U^n$ from the last iteration step should be used:
\begin{equation}
  \begin{aligned}
	\hat{Q}_{2,i,j}^n=& \rho\sigma v_j\tilde{S}_i\frac{U_{2,i+1,j+1}^{n}-U_{2,i-1,j+1}^{n}+U_{2,i-1,j-1}^{n}-U_{2,i+1,j-1}^{n}}{4\Delta\tilde{S}\Delta v}+\frac{1}{2}\sigma^2v_j\frac{U_{2,i,j+1}^{n}+U_{2,i,j-1}^{n}}{(\Delta v)^2}\\
&+\kappa(\theta-v_j)^+\frac{U_{2,i,j+1}^{n}}{\Delta v}-\kappa(\theta-v_j)^-\frac{U_{2,i,j-1}^n}{\Delta v},\quad i=1, ..., I-1, \quad j=1, ..., J-1.
  \end{aligned}
  \label{eq:Q2discrete}
\end{equation}

\section{Boundary conditions}
\label{sec:bcs}
Boundary conditions \eqref{eq:HestonBCfirst}-\eqref{eq:HestonBClast} from \cite{Heston1993} have been widely used in both academia and industry. For the conditions imposed at infinity, it is common to use them on a chosen boundary to find the solution and then truncate the domain into a smaller one to guarantee accuracy, which may cause large computational cost.

 For the boundary condition in the $\tilde{S}$-direction, we use the original boundary condition for $U$ in \cite{Heston1993} to give the boundary condition for $U_2$ as is shown in \eqref{eq:PDEpart2} (we denote it as OriginalBC). Besides, we derive another two boundary conditions for $U_2$ based on the artificial boundary methods in \cite{li2019artificial} which have been proven to be able to improve the accuracy while lower the cost.

We introduce $\tilde{S}_{max}>1$ as an artificial boundary. Then with the artificial boundary methods in \cite{li2019artificial}, two artificial boundary conditions on $\{\tilde{S}=\tilde{S}_{max}\}$ called MApABC1 and MApABC2 can be derived and presented in the same form:
\begin{equation}
\begin{aligned}
&\left(U_{2,\tilde{S}}-\frac{1}{2\tilde{S}_{max}}U_2\right)\bigg|_{\tilde{S}
=\tilde{S}_{max}}\\
=&-\frac{1}{\tilde{S}_{max}}\sqrt{\frac{v}{2\pi}}\int_{0}^{\tau}\left[\frac{2}{v}U_{2,\tau}(\tilde{S}_{max},v,s)+\frac{1}{4}U_2(\tilde{S}_{max},v,s)\right]\exp{\left(-\frac{1}{8}v(\tau-s)\right)}\frac{1}{\sqrt{\tau-s}}ds+ H(v,\tau).
\end{aligned}
\label{eq:ABCcpro}
\end{equation}
In MApABC1,
\begin{equation}
H(v,\tau)=\frac{1}{\tilde{S}_{max}}\int_{0}^{\tau}\left[N\left(\frac{1}{2}\sqrt{v(\tau-s)}\right)-1+\sqrt{\frac{2}{\pi v(\tau-s)}}\exp{\left(-\frac{1}{8}v(\tau-s)\right)}\right]Q(\tilde{S}_{max},v,s)ds;
\label{eq:ABCc1l1}
\end{equation}
In MApABC2,
\begin{equation}
\begin{aligned}
H(v,\tau)=\frac{1}{\tilde{S}_{max}}\int_{0}^{\tau}\int_{\tilde{S}_{max}}^{+\infty}\sqrt{\frac{2}{\pi v(\tau-s)}}\frac{\ln{S^\prime}-\ln{\tilde{S}_{max}}}{v(\tau-s)}\exp{\left(-\frac{\left(\ln{S^\prime}-\ln{\tilde{S}_{max}}+\frac{1}{2}v(\tau-s)\right)^2}{2v(\tau-s)}\right)}Q(S^\prime,v,s)\frac{dS^\prime}{S^\prime} ds.
\end{aligned}
\label{eq:ABCc2l2}
\end{equation}

In addition, we replace the boundary condition \eqref{eq:HestonBClast} for $U$ on $v\to +\infty$ by the one in \cite{Diamond} for accuracy and then for $U_2$, the boundary condition should be of the same form:
 \begin{equation}
U_{2,v}\to 0 \quad \text{as} \quad v \to +\infty,\quad\tilde{S}\in[0,+\infty),\quad \tau\in(0,T].
\label{eq:U0boundaryU2}
\end{equation}
 We also choose $v_{max}$ to restrict the option pricing problem on $[0,\tilde{S}_{max}]\times[0,v_{max}]\times[0,T]$.

The discrete boundary condition on $\tilde{S}=0$ can be given easily:
\begin{equation}
U^n_{2,0,j}=0, \quad j=1, ..., J-1.\\
\end{equation}
For OriginalBC, we have
\begin{equation}
U^n_{2,I,j}=U^n_{2,I-1,j},\quad j=1, ..., J-1.
\end{equation}
For MApABC1 and MApABC2, we have
\begin{equation}
\left(\alpha_j+1-\frac{\Delta\tilde{S}}{2\tilde{S}_{max}}\right)U_{I,j}^{n}-U_{I-1,j}^{n}=\sum_{k=1}^{n-1}\beta_j^{n-k} U_{I,j}^{k}+H_j^n, \quad j=1,..,J-1,
\end{equation}
where
\[\alpha_j=\xi_j+\eta_j,\quad \beta_j^k=\eta_j\phi_j^{k-1}-\alpha_j\phi_j^k,\]
\[\xi_j = \frac{1}{4\sqrt{2\pi}}\frac{\Delta\tilde{S}}{\tilde{S}_{max}}\sqrt{v_j\Delta\tau},\quad \eta_j = \frac{2}{\sqrt{2\pi}}\frac{\Delta\tilde{S}}{\tilde{S}_{max}}\frac{1}{\sqrt{v_j\Delta\tau}},\]
\[\phi_j^0=1, \quad \phi_j^1= \frac{3}{2}\exp{\left(-\frac{1}{8}v_j\tau_1\right)}, \quad \phi_j^k = \frac{1}{\sqrt{k}}\exp{\left(-\frac{1}{8}v_j\tau_k\right)}, \quad k=2, ...,N.\]

For MApABC1, $H_j^n$ can be calculated approximately with the trapezoidal rule applied to \eqref{eq:ABCc1l1} with a discrete $Q$ the same as \eqref{eq:Qdiscrete}. Also, for a better algorithm, we are not supposed to calculate $Q$ in \eqref{eq:ABCc1l1} by $U$ totally from the last iteration step. Instead, we need to divide $Q$  into $Q_1$ and $Q_2$ and use a mixed method for $Q_2$ as  we have done to the equation before.

For MApABC2, We use a family of curves
\[\left(\gamma_0+\gamma_1\ln{\tilde{S}}\right)\exp{\left(-\frac{(\ln{\tilde{S}}-\mu)^2}{2\sigma^2}\right)}\]
where $\gamma_0,\gamma_1,\mu,\sigma$ are all parameters to fit $Q_{j}^n(\tilde{S})$ with $Q_{j}^n(\tilde{S_i})=Q_{i,j}^n,\ i=1, ..., I-1,\ j=1, ..., J-1$ approximated in the same way of \eqref{eq:Qdiscrete}. When the parameters of the curve are determined, the integral in \eqref{eq:ABCc2l2} in the $\tilde{S}$-direction can be easily calculated and a trapezoidal rule is also used in the $\tau$-direction to obtain $H_j^n$. The convergence of the iterative process should be taken into account and dealt with as before.

The boundary conditions in the $v$-direction can be discretized similarly so we ignore the details here.

\section{Numerical experiments}\label{sec:num}

In this section, we show the results of some numerical experiments for the option pricing problem. We implement the iterative splitting method in Section \ref{sec:itsp} with the boundary conditions in Section \ref{sec:bcs}.

Let $\Delta\tau = \Delta\tilde{S} = \Delta v:=h$, we focus on the behaviors of the proposed method using different step sizes. We choose $\tilde{S}_{max}=4$ and $v_{max}=4$ as the artificial boundary of the computational domain. 
The 2nd-order asymptotic solution in \cite{li2019artificial} is used as the reference solution and we compute the relative errors of the numerical results in $l^2$-norm. As for the stopping condition of the iterative process in the iterative splitting method, we set a limitation of $10^{-4}$ on the $l^2$-norm of the  difference between the $U_2$'s calculated in the two adjacent iteration steps.

\begin{figure}[htbp!]
	\centering
	\subfigure{
		\includegraphics[width=0.4\linewidth]{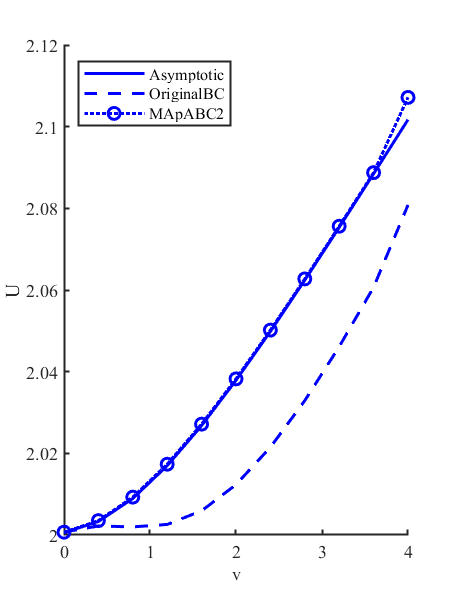}
	}
	\subfigure{
		\includegraphics[width=0.4\linewidth]{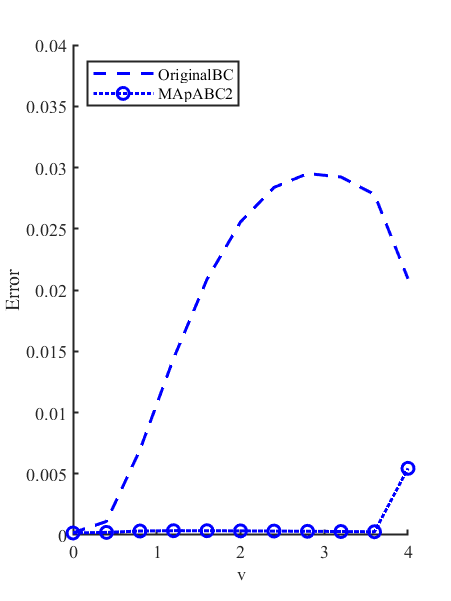}
	}
	\caption{Numerical solutions and errors using the iterative splitting method with OriginalBC and MApABC2 on $\tilde{S}=3$ compared with the reference solution.}
	\label{graph:Sfixed3}
\end{figure}

\begin{figure}[htbp!]
	\centering
	\subfigure{
		\includegraphics[width=0.4\linewidth]{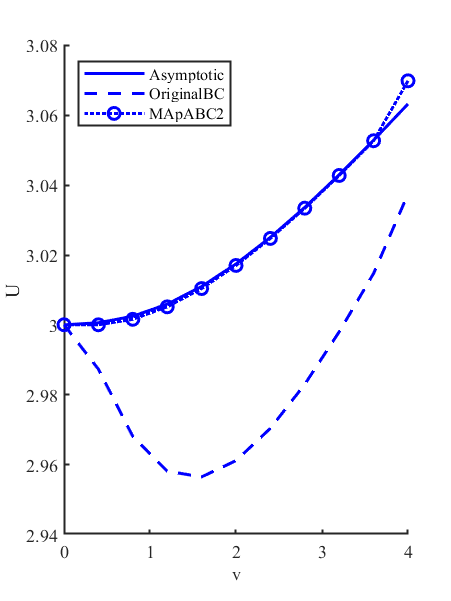}
	}
	\subfigure{
		\includegraphics[width=0.4\linewidth]{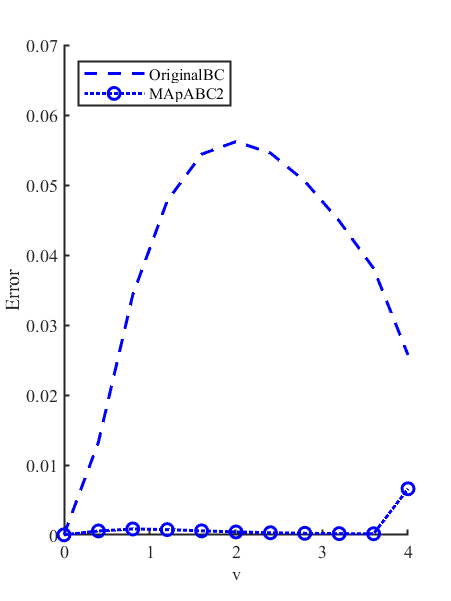}
	}
	\caption{Numerical solutions and errors using the iterative splitting method with OriginalBC and MApABC2 on $\tilde{S}=4$ compared with the reference solution.}
	\label{graph:Sfixed4}
\end{figure}

\begin{figure}[htbp!]
	\centering
	\subfigure{
		\includegraphics[width=0.4\linewidth]{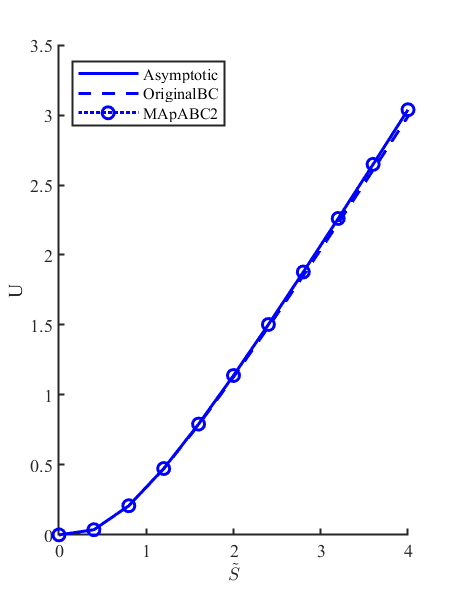}
	}
	\subfigure{
		\includegraphics[width=0.4\linewidth]{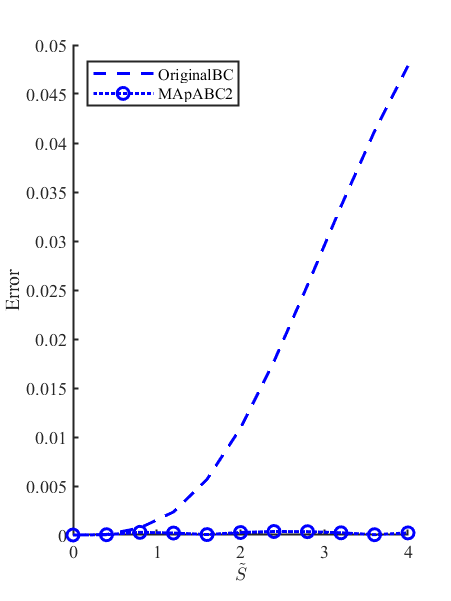}
	}
	\caption{Numerical solutions and errors using the iterative splitting method with OriginalBC and MApABC2 on $v=3$ compared with the reference solution.}
	\label{graph:vfixed3}
\end{figure}

\begin{figure}[htbp!]
	\centering
	\subfigure{
		\includegraphics[width=0.4\linewidth]{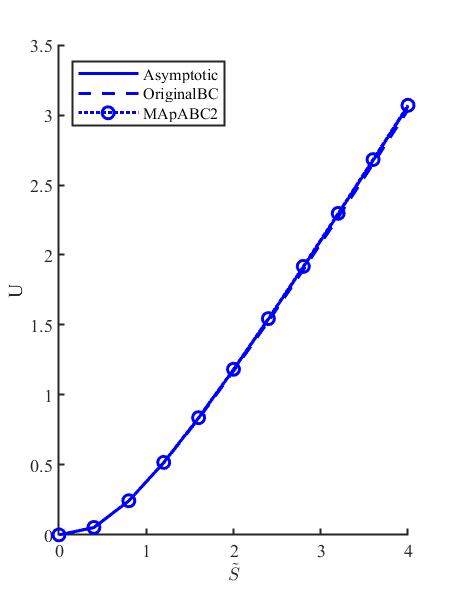}
	}
	\subfigure{
		\includegraphics[width=0.4\linewidth]{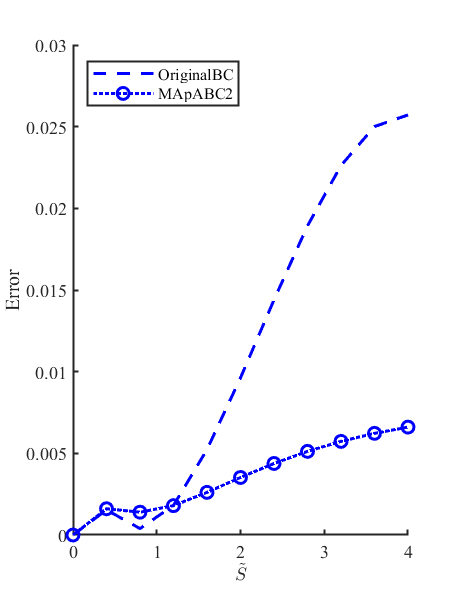}
	}
	\caption{Numerical solutions and errors using the iterative splitting method with OriginalBC and MApABC2 on $v=4$ compared with the reference solution.}
	\label{graph:vfixed4}
\end{figure}

Table \ref{tabl:Parameters-Ex1} shows the first set of parameters  with a large mean reversion rate $\kappa$. The relative errors with respect to the reference solution are presented in Table \ref{tabl:Errors-Ex1}. As the step size gets smaller, the relative error of the solution by the iterative splitting method with MApABC2 decreases and the algorithm achieves almost 1st order convergence. With $h=0.1$ fixed, we show the numerical solutions and the errors with OriginalBC and MApABC2 on $\tilde{S}=3$ and $\tilde{S}=4$ in Figure \ref{graph:Sfixed3} and Figure \ref{graph:Sfixed4} respectively. The numerical solutions and the errors on $v=3$ and $v=4$ are presented in Figure \ref{graph:vfixed3} and Figure \ref{graph:vfixed4} as well. Compared with OriginalBC, MApABC2 gives much better solutions. 

	\begin{table}[h]
		\centering
		\caption{Parameters}
		\label{tabl:Parameters-Ex1}
		\begin{tabular}{p{1.8cm}<{\centering}p{0.6cm}<{\centering}p{0.6cm}<{\centering}p{0.6cm}<{\centering}p{0.6cm}<{\centering}p{0.6cm}<{\centering}}
			\hline
			Parameter & $\kappa$ & $\theta$ & $\sigma$ & $\rho$ & $T$ \\
			\hline
			Value & 5 & 0.08 & 0.1 & -0.6 & 2 \\
			\hline
		\end{tabular}
	\end{table}

\begin{table}[h]
	\centering
	\caption{Relative errors}
	\label{tabl:Errors-Ex1}
	\begin{tabular}{cccccc}
		\hline
		Step size $h$ & 0.4 & 0.2 & 0.1 & 0.05\\
		\hline
		Iterative splitting-OriginalBC & 0.01051 & 0.01174 & 0.01189 & 0.01201 \\
		Iterative splitting-MApABC2 & 0.00407 & 0.00143 & 0.00047 & 0.00020 \\
		\hline
	\end{tabular}
\end{table}

The second set of parameters is shown in Table \ref{tabl:Parameters-Ex2}. We compare the iterative splitting method proposed in Section \ref{sec:itsp} with a classic 2-dimensional finite difference method as in \cite{li2019artificial}. Table \ref{tabl:Errors-Ex2} shows the relative errors of the solutions corresponding to the two methods and boundary conditions with respect to the reference solution.In this example, the mean reversion rate $\kappa$ and the volatility of the variance $\sigma$ are both relatively small, so that the dynamic of the volatility $v(t)$ should be close to a constant. With the iterative splitting method, the first part of the solution $U_1$ accounts for most of the option price $U$ and since we have got the exact solution of $U_1$, a small step size is enough to give a solution of appreciable accuracy.  As is shown in Table \ref{tabl:Errors-Ex2}, the iterative splitting method has a significant advantage over the 2-dimensional finite difference method.
	
	\begin{table}[ht]
		\centering
		\caption{Parameters}
		\label{tabl:Parameters-Ex2}
		\begin{tabular}{p{1.8cm}<{\centering}p{0.6cm}<{\centering}p{0.6cm}<{\centering}p{0.6cm}<{\centering}p{0.6cm}<{\centering}p{0.6cm}<{\centering}}
			\hline
			Parameter & $\kappa$ & $\theta$ & $\sigma$ & $\rho$ & $T$ \\
			\hline
			Value & 0.003 & 0.5 & 0.02 & 0.2 & 2 \\
			\hline
		\end{tabular}
	\end{table}

	\begin{table}[ht]
		\centering
		\caption{Relative errors}
		\label{tabl:Errors-Ex2}\vspace{-2mm}
		\begin{tabular}{ccccc}
			\hline
			Step size $h$ & 0.4 & 0.2 & 0.1 & 0.05\\
			\hline
			 2d finite difference-OriginalBC & 0.04037 & 0.03763 & 0.03653 & 0.03613 \\
     		 2d finite difference-MApABC2 & 0.00788 & 0.00281 & 0.00096 & 0.00048 \\
			\hline
			Iterative splitting-OriginalBC & 0.00012 & 0.00048 & 0.00025 & 0.00013\\
			Iterative splitting-MApABC2 & 0.00011 & 0.00048 & 0.00025 & 0.00013\\
			\hline

		\end{tabular}
	\end{table}
	
In Table \ref{tabl:Parameters-Ex3}, we present the third set of parameters of moderate sizes. We conduct numerical experiments using both the iterative splitting method and the 2-dimensional finite difference method as we  mentioned before and show the relative errors in Table \ref{tabl:Errors-Ex3}. The iterative splitting method with MApABC2 gives the most accurate solution, especially when we set the step size to be $0.05$, the relative error is only $1.6\times 10^{-4}$.
	
	\begin{table}[ht]
  	  \centering
  	  \caption{Parameters}
  	  \label{tabl:Parameters-Ex3}
  	  \begin{tabular}{p{1.8cm}<{\centering}p{0.6cm}<{\centering}p{0.6cm}<{\centering}p{0.6cm}<{\centering}p{0.6cm}<{\centering}p{0.6cm}<{\centering}}
  	  	\hline
  	  	Parameter & $\kappa$ & $\theta$ & $\sigma$ & $\rho$ & $T$ \\
  	  	\hline
  	  	Value & 3 & 0.2 & 0.06 & -0.3 & 2 \\
  	  	\hline
  	  \end{tabular}
	\end{table}
	\begin{table}[ht]
		\centering
		\caption{Relative errors}
		\label{tabl:Errors-Ex3}
		\begin{tabular}{ccccc}
			\hline
			Step size $h$ & 0.4 & 0.2 & 0.1 & 0.05\\
			\hline
			 2d finite difference-OriginalBC & 0.01842 & 0.01558 & 0.01465 & 0.01424 \\
     		 2d finite difference-MApABC2 & 0.00421 & 0.00150 & 0.00070 & 0.00038 \\
			\hline
			Iterative splitting-OriginalBC & 0.00746 & 0.00815 & 0.00857 & 0.00878\\
			Iterative splitting-MApABC2 & 0.00265 & 0.00084 & 0.00031 & 0.00016\\
			\hline
		\end{tabular}
	\end{table}

\begin{figure}[htbp!]
	\centering
	\subfigure{
		\includegraphics[width=0.4\linewidth]{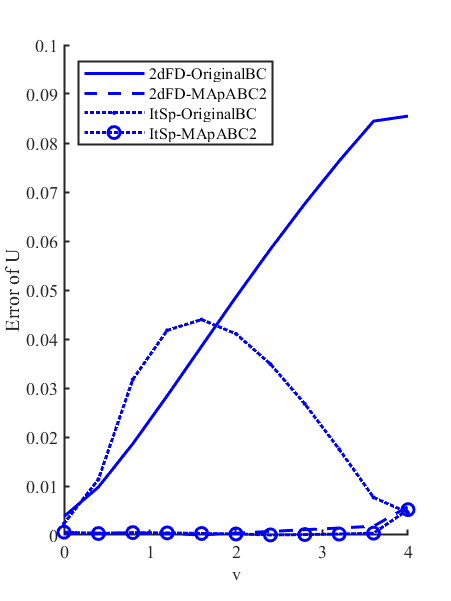}
	}
	\subfigure{
		\includegraphics[width=0.4\linewidth]{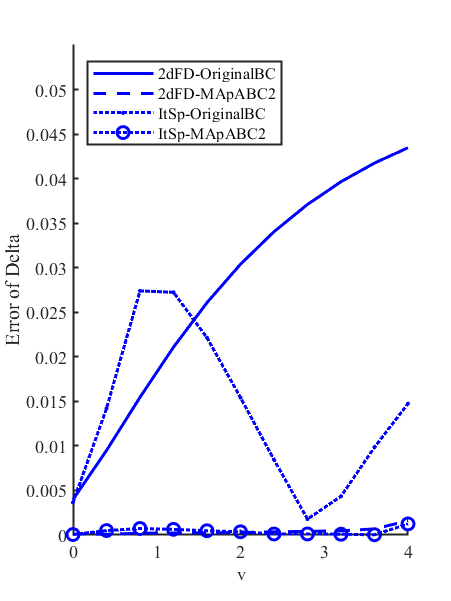}
	}
		\subfigure{
		\includegraphics[width=0.4\linewidth]{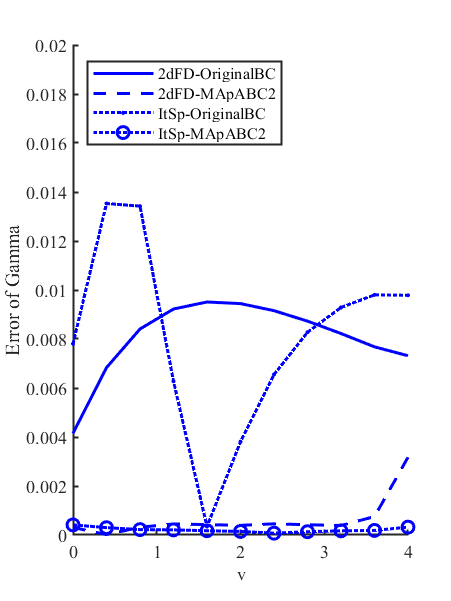}
	}
	\subfigure{
		\includegraphics[width=0.4\linewidth]{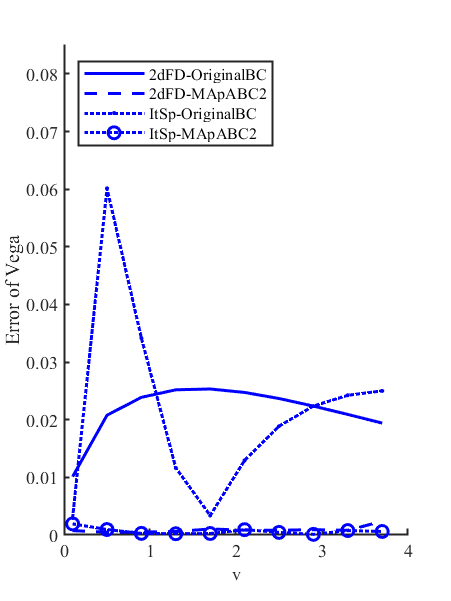}
	}
	\caption{Errors of the option price, Delta, Gamma and Vega using the 2-dimensional finite difference scheme  and the iterative splitting method (represented as '2dFD' and 'ItSp' respectively) with OriginalBC and MApABC2  on $\tilde{S}=4$.}
	\label{graph:Greeks}
\end{figure}

Some of the derivatives of the option price with respect to different variables are called Greeks in finance and are of high importance in industry. For example, the first and second derivatives of the option price with respect to the underlying price are called Delta and Gamma, while Vega corresponds to the first derivative with respect to the variance of the underlying. Greeks are good measures of risk and sometimes determine the trading volume so the computational efficiency and accuracy have received wide attention. With $h=0.1$ fixed, we plot the errors of the option price and Greeks on $\tilde{S}=4$ in Figure \ref{graph:Greeks}. The numerical results show that the iterative splitting method with MApABC2 behaves the best in terms of accuracy, and the calculation is simpler and more convenient as well since the iterative splitting method helps to transform a 2-dimensional problem into two quasi 1-dimensional ones.
	
\section{Conclusions}\label{sec:con}
We propose an iterative splitting method to solve the option pricing problem and concentrate on the Heston stochastic volatility model. We introduce the artificial boundary conditions  based on the methods from \cite{li2019artificial}. Numerical study has shown that for the European options, the iterative splitting method can improve the accuracy by contrast with the classic two-dimensional finite difference scheme. Using the artificial boundary condition called MApABC2, the solution and the Greeks are calculated better in terms of the error. The iterative splitting method transformed a two-dimensional PDE into two quasi one-dimensional PDEs with the idea of operator splitting. The convergence of the iteration is ensured by a mixed method. Our new method is easy to interpret, understand and implement. It can also be extended to other models and options, and then the existed methods for the Black-Scholes model can be available, which will be considered in the future.

\bibliographystyle{siam}
\bibliography{bib}

\end{document}